\documentstyle[epsfig,prb,aps]{revtex}
\begin{document}

\wideabs{

\title{Spectroscopic Studies of the Vibrational and Electronic
Properties of Solid Hydrogen to 285 GPa}

\author{Alexander F. Goncharov, Eugene Gregoryanz, Russell
J. Hemley, Ho-kwang Mao}

\address{Geophysical Laboratory and Center for High Pressure
Research, Carnegie Institution of Washington, \\
5251 Broad Branch Road NW, Washington D.C. 20015 U.S.A}


\maketitle

\begin{abstract}

We report Raman scattering and visible to near-infrared absorption spectra of
solid hydrogen under static pressure up to 285 GPa at 85-140 K.
We obtain pressure dependences of vibron and phonon
modes in agreement with previously determined to lower pressures.
The results indicate the stability of the ordered molecular phase
III to the highest pressure reached and provide constraints on the
insulator-to-metal transition pressure.

\end{abstract}
 }

Recent theoretical predictions for transformation of crystalline
solid hydrogen to its metal state are still uncertain and range
between 260-410 GPa \cite{nagao,natoli,johnson}. Metallization by
band overlap is predicted to occur prior break down to a
monoatomic state \cite{johnson}. The ability of theory to
calculate the band gap at relevant densities is substantially
impaired by the uncertainty of the structure of the high-pressure
molecular modification of solid hydrogen -phase III (e.g., Ref.
\cite{RMP}). The nature of this phase and its related infrared
activity have been the topic of numerous theoretical and
experimental studies (e.g. Ref. \cite{kohanoff}), but definitive
knowledge of its crystal and electronic structure is not yet in
hand.

Diamond-anvil cells have been used successfully to reach static
pressures on the order of 360 GPa for compressed metals
\cite{vohra90,mao90,hemley97}.  However, numerous attempts to
compress solid hydrogen in order to transform it to conducting
states \cite{mao89,eggert90,mao,eggert91,chen,hemley96,ruoff} have
not been successful in reaching the critical pressure range while
at the same time characterizing the sample and pressure
definitively. The claim of compressing solid hydrogen to 342 GPa
\cite{ruoff}, for instance, showed no evidence for the presence of
hydrogen in the sample chamber, indicating that instead the "soft"
hydrogen was most likely lost by developing small leaks in
diamonds and gasket or by reaction with the gasket material (see
Ref. \cite{RMP}). In addition, the large stress-induced increase
in optical absorption and fluorescence in diamond anvils
\cite{vohra90,mao91,ruoff91} poses a major obstacle in optical
measurements of hydrogen samples and pressure calibration by ruby
fluorescence. Control measurements for optical absorption
experiments are essential. For example, comparison of the
absorption of the transparent ruby grains adjacent to hydrogen can
be used to ascertain that the pressure-induced changes in
absorption occurs is in hydrogen but not in the diamond windows
\cite{mao89}. At pressures beyond 180 GPa, the ruby fluorescence
becomes extremely weak and can be overwhelmed by diamond
fluorescence, thus presenting another serious problem for pressure
calibration.  As a result, the pressure for our previously
reported onset of absorption in hydrogen could only be determined
as being above 200 GPa, but the upper limit could not be
established. Indirect methods of pressure calibration such as
x-ray diffraction measurements on the gasket \cite{ruoff} do not
indicate the pressure of the sample, and pressure calibration
based on the pressure shift of diamond Raman band
\cite{hanfland85} may vary by as much as a factor of three
depending upon the local nonhydrostatic stress condition on the
pressure-bearing diamond anvil surface \cite{schiferl97,xu00}.

Infrared and Raman spectroscopies have been successfully used to
obtain information about molecular orientational ordering,
strength of intermolecular interactions, crystal structure, phase
transitions, and charge transfer in hydrogen, but have been
previously limited in the pressure range reached
\cite{chen,mazin,hanfland,hanfland1,hanfland2,goncharov,hemley88,lorenzana}.
In this paper we extend spectroscopic measurements on solid
hydrogen to 285 GPa, as well as accurate pressure determination by
ruby fluorescence to 255 GPa. IR spectroscopy is particularly
suited for ultrahigh pressure study of hydrogen because of the
dramatic increase in vibron intensity in phase III. On the other
hand, one must overcome the difficulty of focusing the diffraction
limited IR beam to study microscopic samples in the diamond cell.
For Raman spectroscopy, we must reduce the fluorescence background
of the diamond by choosing anvils with very low initial
fluorescence. We find that the hydrogen vibron persists to the
highest pressure reached, indicating that the hydrogen molecules
remain intact.  We find that phase III persists while no
unambiguous change of optical properties of hydrogen could be
detected. We constrain the pressure of the transition to the
metallic state to 325-495 GPa.

Our Raman and IR techniques are described elsewhere
\cite{goncharov,hemley98,hemley98a,goncharov99}. Near IR
measurements were performed with a conventional and synchrotron
source. We used natural type I beveled (8-10 degrees) diamonds
with 30-50 $\mu$m culets. Here, we report the results of four
different experiments, which ended up by diamond failure at
230-285 GPa. All experiments were done at low temperatures (78-140
K). In one of them (to 255 GPa) the sample contained a small
amount of ruby, so Raman and IR absorption measurements could be
performed along with accurate pressure measurements by ruby
fluorescence using a direct pumping scheme with a Ti-sapphire
laser (705-740 nm) as described in Ref. \cite{goncharov99}.  Raman
measurements were also performed with near-IR excitation.  Other
samples contained a large amount of ruby to improve the pressure
distribution in the high-pressure chamber. Since the vibron
absorbance in phase III is rather high \cite{hanfland}, we could
afford to reduce an amount of hydrogen to approximately 1\% of the
whole sample volume at highest pressures. Visible optical
absorption spectroscopy was used to further characterize the
samples.

\begin{figure} \centerline{\epsfig{file=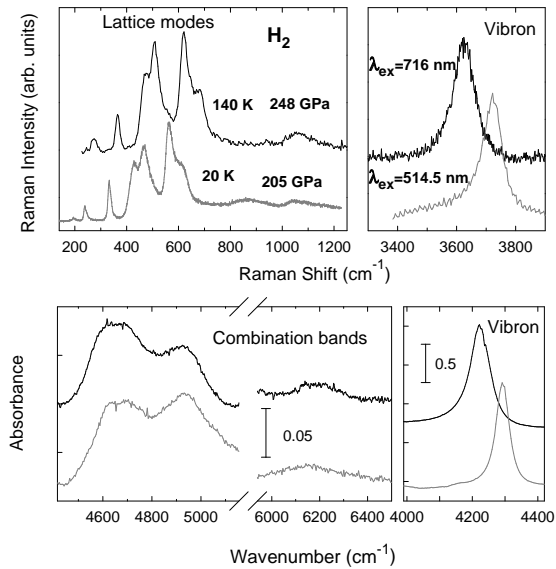,width=10cm}}
\caption{Raman and IR absorption spectra of H$_2$ at 248 GPa, 140 K
(upper curves) and 205 GPa, 20 K (lower curves).} \label{fig1} \end{figure}

Figure \ref{fig1} presents Raman and IR spectra at 250 GPa and at
205 GPa ({\it p}-H$_2$ \cite{goncharov}). The latter were obtained
on a sample converted to pure $p$-H$_2$ \cite{goncharov} Raman
spectra at 250 GPa reveal of a number of narrow low-frequency
bands associated with librations and translations of molecules in
accord with our previous study to 205 GPa in {\it p}-H$_2$
\cite{goncharov}. The spectra are qualitatively very similar; the
pressure shift of the lattice and vibron modes will be discussed
below.  As in Refs. \cite{hemley88,lorenzana}, we observe only one
Raman vibron. The second one (which is much weaker) observed in
Ref. \cite{goncharov} is outside our available spectral range
determined by the excitation wavelength and the sensitivity of the
CCD detector. IR spectra are also in agreement with previous
studies at lower pressures
\cite{chen,hanfland,hanfland1,hanfland2,goncharov}. Comparison of
the frequencies of the high-frequency IR bands with those of the
low-frequency Raman spectrum suggest that they originate from
combinations of the IR vibron and lattice modes
\cite{hanfland,hanfland1,hanfland2}.

\begin{figure} \centerline{\epsfig{file=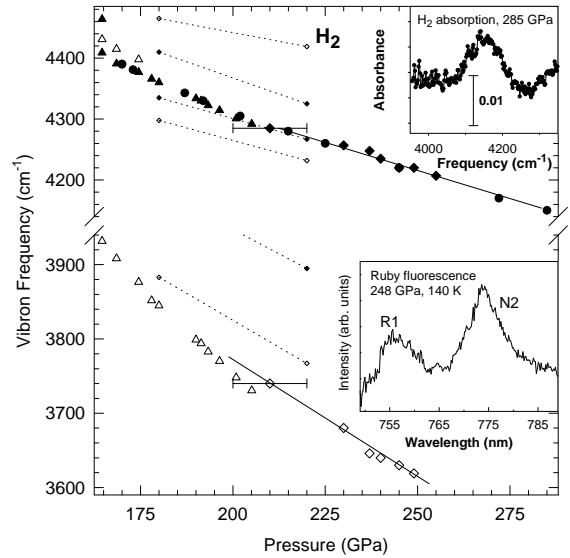,width=10cm}}
\caption{Pressure dependence of the measured Raman and IR vibron
frequencies compared to theoretical results. The large solid and
open symbols are our IR and Raman data respectively. The solid curves
are straight lines (guide to the eye). The small solid and open symbols
connected by dotted lines are the calculations from the Ref.
\protect\cite{kohanoff99}. The inset shows the IR absorption
spectrum of composite a H$_2$/ruby sample at 285 GPa.  The inset in
the bottom is the ruby fluorescence spectrum at 248 GPa measured
by the direct pumping technique with 705 nm excitation.  The pressure
is determined from the spectral positions of N$_2$ and R$_1$ ruby
lines according to the calibration of Ref.  \protect\cite{eggert}
extrapolated to higher pressures. The accuracy of the pressure
determination is estimated as $\pm$5\%.} \label{fig2} \end{figure}

\noindent Of particular interest is the 6200 cm$^{-1}$ band, which
represents the combination of the IR vibron and the highest
frequency lattice mode, corresponding to the axial translational
vibration \cite{hanfland,hanfland1,hanfland2}.

\begin{figure} \centerline{\epsfig{file=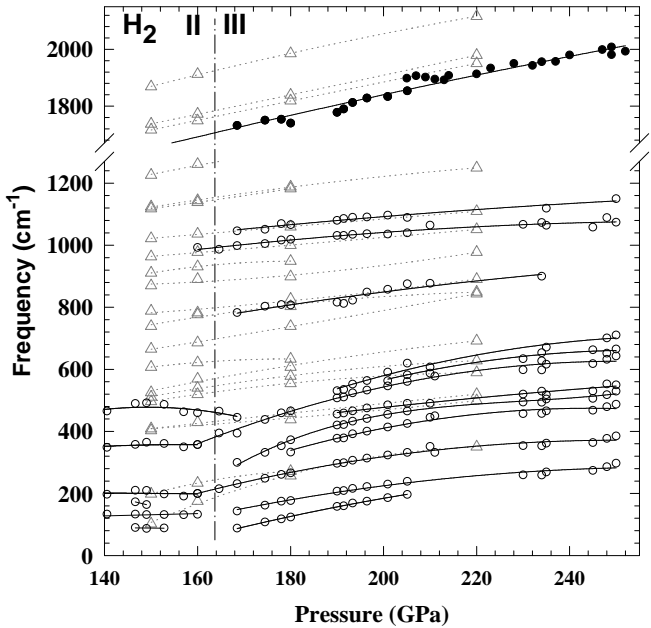,width=10cm}}
\caption{Pressure dependence of the measured lattice mode
frequencies and comparison with theoretical results.
The solid and open symbols are our IR and Raman data,
respectively. The solid lines are guides to the
eye. The triangles connected by dotted lines are calculations from
the Ref. \protect\cite{kohanoff99}. The pressure determination is
described in the text and caption to Fig. \protect\ref{fig2}
.}
\label{fig3} \end{figure}

Figure \ref{fig2} shows the pressure dependence of the IR- and
Raman-active
 vibrons and lattice mode frequencies. The results obtained are
 in reasonable agreement with our previous study
 \cite{goncharov}. In contrast, the pressures determined
 in Refs. \cite{hanfland,hanfland2} by extrapolating the IR vibron
 frequency by the second order polynomial are substantially
 underestimated as noted in Ref. \cite{charge}. For the
 IR absorption measurements above 250 GPa (Fig. \ref{fig2} inset),
 the pressure was determined by linearly extrapolating the
 IR frequency. This determines the stability range of phase
 III on the basis optical measurements. Comparison of the
 measured frequencies with the results of recent theoretical
 calculations \cite{kohanoff99} shows some discrepancies
 including both the number of the modes observed and their
 IR and Raman activity, however the pressure dependence (slope) is
 in reasonable agreement (Fig. \ref{fig2}). The same
 is true for the lattice modes (Fig. \ref{fig3}), although
 direct comparison is complicated because of a lack of specific
 assignment from both experiment and theory.

\begin{figure} \centerline{\epsfig{file=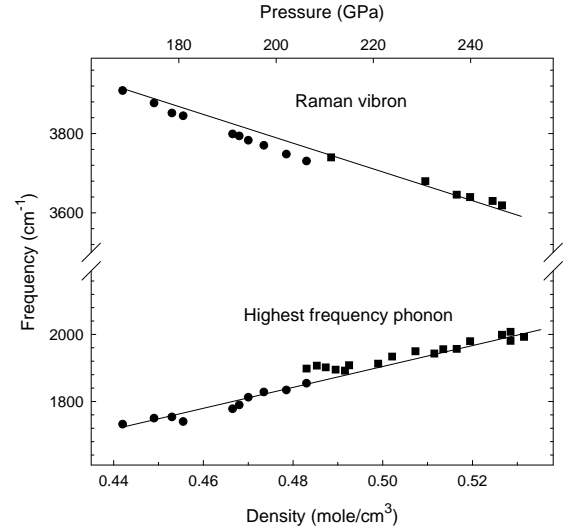,width=10cm}}
\caption{Raman vibron frequency and highest phonon frequency
(see text) as a function of density calculated according
to measurements from the Ref. \protect\cite{loubeyre} and
extrapolated to high pressures. } \label{fig4} \end{figure}

The reason for the disagreement in the case of the vibron modes
seems to originate from the structure of phase II proposed in Ref.
\cite{kohanoff99} (very similar structure was proposed earlier
\cite{tse}), which contains two different site symmetries for
hydrogen molecules (with generally different bond length).  As a
result, two distinct manifolds of vibrons are predicted, which does
not seem to be confirmed by the experiment. The experimental data
indicate that the splitting between IR and Raman vibrons is
dominated by vibrational (or factor-group) splitting
\cite{hanfland}, and the multiple site-group splitting (if any)
does not play a substantial role, unlike the situation for several other
molecular crystals (e.g., $\delta$-N$_2$ \cite{lesar}). If so, one
can use a simplified bond-force model to predict the upper bound
for dissociation of the hydrogen molecules (e. g. Ref.
\cite{ashcroft90}). The distinction between intra- and
intermolecular bonding creates a gap in the vibrational density of
states, which can be determined as a difference between the lowest
vibron and the uppermost phonon (Fig. \ref{fig4}). Linear
extrapolation of the frequencies to high density (pressure) give an
intersection at $\sim$0.74 mole/cm$^{-3}$ (495 GPa if using EOS
from Ref. \cite{loubeyre}) for the point corresponding to the "no
gap" situation that is to the transition to the nonmolecular state
(presumably metallic) assuming no further phase transitions occur.
This kind of transformation path would require a structural
resemblance between the nonmolecular and molecular phase (e.g,
ice X \cite{goncharov96}) and may be
preempted by band-overlap metallization within the molecular state,
which could trigger a transformation to an altogether different molecular structure \cite{johnson,mazin95}.

\begin{figure} \centerline{\epsfig{file=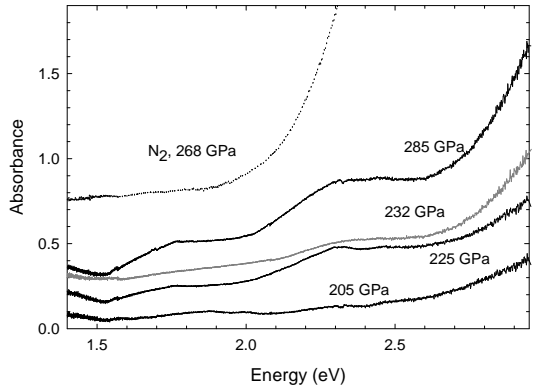,width=10cm}}
\caption{Optical absorption spectra of a composite H$_2$/ruby
sample in a diamond anvil cell at different pressures and
comparison with a similarly prepared N$_2$ sample
\protect\cite{gregoryanz}. The solid black and gray lines
correspond to experiments with different H$_2$ content. }
\label{fig5} \end{figure}

Figure \ref{fig5} shows optical absorption spectra of hydrogen
samples mixtured with ruby.  With increasing pressure an absorption
edge appears on the high-energy side and shifts to lower energies.
This absorption is presumably related to a closure of the band gap
of diamond under nonhydrostatic conditions, as calculated in Ref.
\cite{nielsen} and observed experimentally
\cite{vohra90,ruoff91,syassen}. This absorption has a long tail
with additional features. The 2.3 eV feature observed above 220
GPa has been previously reported in reflectivity spectra when
metallic gaskets measured through the stressed diamond anvil
\cite{vohra91} and some of our N$_2$ absorption spectra
\cite{gregoryanz}; a peak at this energy can also be observed in
fluorescence spectra of stressed diamonds \cite{mao91}. Comparison
of spectra measured for two samples with different H$_2$ contents do not show substantial
differences. Thus, no obvious absorption due to hydrogen could be
observed in these experiments. For comparison, in our experiments on
N$_2$ under similar conditions the absorption edge due to
electronic interband transitions in the sample is quite obvious (Fig.
\ref{fig4}), although thicker samples without ruby filler were
needed to determine the band gap \cite{gregoryanz}.

Resonance Raman scattering could provide an alternative
independent way of measurement of the band gap since one expects a
resonant increase of intensity of the Raman signal when incoming
(and outgoing) phonon energies approach those of electronic
excitations, including those associated with the band gap
\cite{martin}. We found evidence for this in our early experiments on
H$_2$/ruby composite samples, although the pressure onset and the
measure of the degree of enhancement could not be determined
quantitatively \cite{mao89}. Accurate measurements of this kind
require the use of an internal standard (or
precision positioning of the sample) and correction for the
absorption of hydrogen. The latter is not known because the
background absorption of stressed diamond masks it (see above).
Nevertheless, here we observed an increase of intensity (by a
factor 1.5-2.5) of the librational modes with 716-724 nm
excitation at highest pressures reached (250-255 GPa) in our Raman
experiment. Because of
diamond failure, we were unable to determine whether any further
increase in Raman intensity occurs at higher pressure.

\begin{figure} \centerline{\epsfig{file=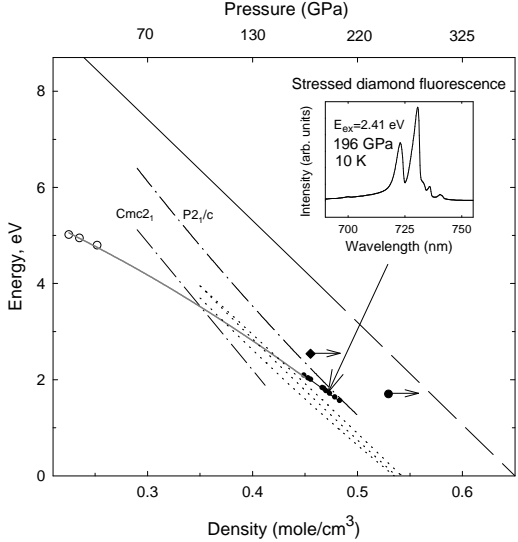,width=10cm}}
\caption{Band gap of hydrogen as a function of density. The solid
line is the best fit to data of Ref. \protect\cite{hemley91}
extended linearly to gap closure (long dashed line). The long
dash-dotted lines  are the theoretical calculations for different
crystal structures of phase III \protect\cite{johnson}. The dotted
lines are the theoretical calculations for Pca2$_1$, Cmc2$_1$ and
P2$_1$/c structures (from left to right, respectively) of phase
III from Ref. \protect\cite{stadele} (the results are very close
to each other). The solid diamond is Raman resonance scattering
point obtained in Ref. \protect\cite{mao89}. The solid circle is
the similar point obtained in this work. The arrows show that the
pressure could be underestimated \protect\cite{mao89} or the
resonance conditions were not reached (see text). For comparison,
estimations of the band gap of stressed diamond (this work) are
also shown (inset shows the stressed- induced fluorescence of
diamond at sample pressure of 196 GPa and 10 K). The double solid
- dotted line corresponds to measurements of the peak positions of
two strongest diamond fluorescence peaks, which may correspond to
the band gap (no correction for the phonon energy, possibly
involved in case of indirect gap, is made). The open circles are
the absorption edge values of stressed type II diamond obtained
from direct measurements under pressure \protect\cite{syassen}.
The gray solid line is a guide to the eye. The hydrogen density is
calculated from Ref. \protect\cite{loubeyre} (extrapolated above
120 GPa) by using the pressure measured inside the diamond anvil
cell for all experimental data.} \label{fig6}
\end{figure}

Figure \ref{fig6} summarizes the results of determination of the
band gap of hydrogen as a function of density by various methods.
The extrapolation of the results of dielectric oscillator model
fits based on interference fringe measurements \cite{hemley91} can
be considered a rough guide for the density dependence of the
direct gap. Recently calculated theoretically band gaps for
assumed Pca2$_1$, P2$_1$/c and Cmc2$_1$ structures
\cite{johnson,stadele} are in agreement in terms of the slope but
show smaller values, which is consistent with the fact that they
are indirect gap values. Notably, some of the early calculations
predicted very different density dependence for the direct and
indirect gaps \cite{RMP,hemley91}. The indications of the band gap
energy from the resonance Raman scattering (Ref. \cite{mao89} and
this work) are also shown and are consistent with the general
trend. No evidence for the direct band gap closure is found in our
visible transmission measurements (Fig. \ref{fig5}), although a
small indirect gap or even its closure cannot be ruled out because
of a submicron thickness of the sample (see also Ref.
\cite{mao89}). It is worth noting that the band gap of the
stressed diamond also decreases with pressure
\cite{nielsen,syassen}. In one of our early experiments with
synthetic diamonds \cite{goncharov} we observed a stressed induced
fluorescence as narrow bands shifting with pressure to lower
energy (Fig. \ref{fig6} inset), which can be related to the band
gap closure of stressed diamond in agreement with data of Ref.
\cite{syassen} and calculations \cite{nielsen}. Note that the band
gap of hydrogen and stressed diamond are estimated to be
comparable at these pressures, which makes measurements of the
band gap of hydrogen problematic by conventional optical
techniques. Nevertheless, the data presented in Fig. \ref{fig6}
show that according to different estimations the direct band gap
of hydrogen is expected to close at about 0.6-0.65 mole/cm$^3$,
which correspond to pressures of 325-385 GPa (c.f. Ref.
\cite{hemley91}) according the equation of state of Ref.
\cite{loubeyre}.

In conclusion, we have performed various optical measurements of
hydrogen to 285 GPa. Vibrational and optical spectroscopy data
provide different sets of constraints on the higher-pressure
transformations, including both pressure induced dissociation to
form a non-molecular metallic solid and band-overlap metallization
in which molecular bonding persists. Extrapolations of the vibron
and phonon frequencies suggest transformation to a monoatomic
state below 495 GPa. On the other hand, considerations of the
absorption edge indicate the pressure of metallization at 325-385
GPa on the basis of tentative extrapolation of the direct band gap
energy. Although this complicated by affects of stressed-induced
diamond absorption and possible differences between the behavior
of the direct and indirect gap, there appears to be an emerging
consistence between various experimental and theoretical results,
with a predicted transition at 325-495 GPa.

We acknowledge financial support of CIW, NSLS, NSF and W. M. Keck
Foundation.




\begin{references}

\bibitem{nagao} Nagao, K., Nagara, H., and Matsubara,
S. (1997) Phys. Rev. B. {\bf 56}, 2295-2298.

\bibitem{natoli} Natoli, V., Martin, R. M., and Ceperley,
D. M. (1993) Phys. Rev. Lett. {\bf 70}, 1952-1955.

\bibitem{johnson} Johnson, K. A. and Ashcroft, N. W. (2000)
Nature {\bf 403}, 632-635.

\bibitem{RMP} Hemley, R. J., Mao, H. K. (1994) Rev. Mod. Phys. {\bf 66}
671-692.

\bibitem{kohanoff} Kohanoff, J. (2001) J. Low
Temp. Phys. {\bf 122}, 297-311.

\bibitem{vohra90} Vohra, Y. K., Luo, H., and Ruoff,
A. L. (1990) Appl. Phys. Lett. {\bf 57}, 1007-1111.

\bibitem{mao90} Mao, H. K., Wu, Y., Chen, L. C., Shu, J., and Jephcoat, A. P.
(1990) J. Geophys. Res., {\bf 95}, 21737-21742.

\bibitem{hemley97} Hemley, R. J., Mao, H. K., Shen, G., Badro, J., Gillet, P.,
Hanfland, M., H\"ausermann (1997) Science
{\bf 276}, 1242-1245.

\bibitem{mao89} Mao, H. K. and Hemley, R. J. (1989) Science
{\bf 244}, 1462-1465.

\bibitem{eggert90} Eggert, J. H., Goettel, K. A. and Silvera,
I. F. (1990) Europhys. Lett. {\bf 11}, 775-781; {\bf 12}, 381
(Addendum).

\bibitem{mao} Mao, H. K., Hemley, R. J., and Hanfland,
M. (1990) Phys. Rev. Lett. {\bf 65}, 484-487.

\bibitem{eggert91} Eggert, J. H., Moshary, F., Evans,
W. J., Lorenzana, H. E., Goettel, K. A., Silvera, I. F.,
and Moss, W. C. (1991) Phys. Rev. Lett. {\bf 66}, 193-196.

\bibitem{chen} Chen, N. H., Sterer, E., and Silvera,
I. F. (1996) Phys. Rev. Lett. {\bf 76}, 1663-1666.

\bibitem{hemley96} Hemley, R. J., Mao, H. K., Goncharov,
A. F., Hanfland, M., Struzhkin, V. (1996) Phys. Rev. Lett. {\bf
76}, 1667-1670.

\bibitem{ruoff} Narayana, C., Luo, H., Orloff, J., and
Ruoff, A. L. (1998) Nature {\bf 393}, 46-49.

\bibitem{mao91} Mao, H. K. and Hemley, R. J. (1991) Nature
{\bf 351}, 721-724.

\bibitem{ruoff91} Ruoff, A. L., Luo, H., and Vohra,
Y. K. (1991) J. Appl. Phys. {\bf 69}, 6413-6416.

\bibitem{hanfland85} Hanfland, M. and Syassen, K. (1985)
J. Appl. Phys. {\bf 57}, 2752-2756.

\bibitem{schiferl97} Schiferl, D., Nicol, M., Zaug, J. M.,
Sharma, S. K., Cooney, T. F., Wang, S. Y., Anthony, T. R.,
and Fleischer, J. F. (1997) J. Appl. Phys. {\bf 82}, 3256-3265.

\bibitem{xu00} Xu, J. and Mao, H. K. (2000) Science,
{\bf 290}, 783-785.

\bibitem{mazin} Mazin, I. I., Hemley, R. J., Goncharov,
A. F., Hanfland, M. and Mao, H. K. (1997) Phys. Rev. Lett. {\bf
78} 1066-1069.

\bibitem{hanfland} Hanfland, M., Hemley, R. J., Mao,
H. K. (1993) Phys. Rev. Lett. {\bf 70}, 3760-3763.

\bibitem{hanfland1} Hanfland, M., Hemley, R. J., i
Mao, H. K., and Williams, G. (1992) 
Phys. Rev. Lett. {\bf 69}, 1129-1132.

\bibitem{hanfland2} Hanfland, M., Hemley,
R. J., Mao, H. K. (1994) {\it High-Pressure Science and
Technology}-1993, edited by S. C. Schmidt {\it et al.} (AIP,
New York), p. 877-880.

\bibitem{goncharov} Goncharov, A. F., Hemley, R. J., Mao,
H. K., Shu, J. (1998) Phys. Rev. Lett. {\bf 80}, 101-104.

\bibitem{hemley88} Hemley, R. J., Mao, H. K. (1988)
Phys. Rev. Lett. {\bf 61}, 857-860.

\bibitem{lorenzana} Lorenzana, H. E., Silvera, I. F.,
and Goettel, K. A. (1989) Phys. Rev. Lett. {\bf 63}, 2080-2083.

\bibitem{hemley98} Hemley, R. J., Goncharov, A. F., R. Lu,
Struzhkin, V. V., Mao, H. K. (1998) Il Nuovo Cimento {\bf 20},
539-551.

\bibitem{hemley98a} Hemley, R. J., Goncharov, A. F., Mao, H. K., Karmon, E.,
and Eggert, J. H. (1998) J. Low Temp. Phys. {\bf 110}, 75-88.


\bibitem{goncharov99} Goncharov, A. F., Struzhkin, V. V.,
Hemley, R. J., Mao, H. K., and Liu, Z. (1999) in: {\it Science and
Technology  of High Pressure}, edited by M. H. Manghnani, W.
J. Nellis and M. F. Nicol (Universities Press, Hyderabad,
India, Honolulu, Hawaii, 1999), Vol. 1, p. 90-95.


\bibitem{kohanoff99} Kohanoff, J., Scandolo, S., de Gironcoli, S.,
and Tosatti, E. (1999) Phys. Rev. Lett. {\bf 83}, 4097-4100.

\bibitem{eggert} Eggert, J. H., Moshary, F., Evans,
W. J., Goettel, K. A., and Silvera, I. F., (1991) Phys. Rev. B
{\bf 44}, 7202-7208.

\bibitem{charge} Hemley, R. J., Mazin, I. I.,  
Goncharov, A. F., and Mao, H. K., (1997) Europhys. Lett. {\bf 37}, 403-407.

\bibitem{loubeyre}  Loubeyre P., LeToullec, R., H\"ausermann, D., Hanfland, M.,
Hemley, R. J., Mao, H. K., and Finger, L. W. (1996) Nature {\bf
383}, 702-704.


\bibitem{tse} Tse, J., Klug, D. (1995) Nature {\bf 378}, 595-597.

\bibitem{lesar} LeSar, R., Ekberg, S. A., Jones L. H., Mills, R. L., Schwalbe,
L. A., Schiferl, D. (1979) Sol. State Commun. {\bf 32}, 131-134.

\bibitem{ashcroft90} Ashcroft, N. W. (1990) Phys. Rev. B {\bf 41}, 10963-10971.

\bibitem{goncharov96} Goncharov, A. F., Struzhkin, V. V., Somayazulu, M. S.,
Hemley, R. J., and Mao, H. K. (1996) Science {\bf 273}, 218-220.

\bibitem{mazin95} Mazin, I. I. and Cohen, R. E. (1995) Phys. Rev. B
{\bf 52}, R8597-R8600.

\bibitem {gregoryanz} Gregoryanz, E., Goncharov, A. F.,
Hemley, R. J., and Mao, H. K. (2001) Phys. Rev. B {\bf 64},
052103-052106.

\bibitem{nielsen} Nielsen, O. H. (1986) Phys. Rev. B {\bf 34},
5808-5819.

\bibitem {vohra91} Vohra, Y. K. (1991) in:
{\it Recent Trends in High Pressure Research}
Proc. of the XIII International Conference on 
High Pressure Science and Technology, edited by A. K. Singh. pp. 349-358.

\bibitem{syassen} Syassen, K. (1982) Phys. Rev. B {\bf 25}, 6548-6550.


\bibitem{stadele} St\"adele, M. and Martin, R. (2000) 
Phys. Rev. Lett. {\bf 84}, 6070-6073.

\bibitem {martin} Martin, R. M., and Falicov, L. M. in:{\it
Light Scattering in Solids} Topics in Applied Physics, vol. 8,
edited by M. Cardona (Springer-Verlag, New York 1983), pp. 79-145.

\bibitem{hemley91} Hemley, R. J., Hanfland, M. and Mao, H. K. (1991)
Nature {\bf 350}, 488-491.


\end{references}
\end{document}